\documentclass[12pt,preprint]{aastex}
\usepackage{emulateapj5}
\usepackage{dblfloatfix}
\usepackage{float}
\usepackage{fixltx2e}
\usepackage{epsfig}
\begin{document}
\shorttitle{PUC}
\shortauthors{Kollmeier et al.}

\newcommand{\hmpc}{{\rm Mpc h^{-1}}}
\newcommand{\ghi}{\Gamma_{\rm HI}}
\newcommand{\gh}{$\Gamma_{\rm HI}$}
\newcommand{\Nhi}{$N_{\rm HI}$}
\newcommand{\nhi}{N_{\rm HI}}
\newcommand{\fesc}{f_{\rm esc}}
\newcommand{\msun}{M_{\odot}}
\newcommand{\kms}{\, {\rm km\, s}^{-1}}
\newcommand{\mas}{\, {\rm mas\, yr}^{-1}}
\newcommand{\cm}{\, {\rm cm}}
\newcommand{\gm}{\, {\rm g}}
\newcommand{\erg}{\, {\rm erg}}
\newcommand{\kel}{\, {\rm K}}
\newcommand{\kpc}{\, {\rm kpc}}
\newcommand{\mpc}{\, {\rm Mpc}}
\newcommand{\seg}{\, {\rm s}}
\newcommand{\kev}{\, {\rm keV}}
\newcommand{\hz}{\, {\rm Hz}}
\newcommand{\etal}{et al.\ }
\newcommand{\yr}{\, {\rm yr}}
\newcommand{\mpyr}{{\rm mas}\, {\rm yr}^{-1}}
\newcommand{\gyr}{\, {\rm Gyr}}
\newcommand{\eq}{eq.\ }
\newcommand{\lya}{Lyman-$\alpha$}

\def\arcsec{''\hskip-3pt .}
\newcommand{\bdv}[1]{\mbox{\boldmath$#1$}}
\title{\bf The Photon Underproduction Crisis}

\author{Juna A. Kollmeier\altaffilmark{1}, David H. Weinberg\altaffilmark{2}, Benjamin D. Oppenheimer\altaffilmark{3}, Francesco Haardt\altaffilmark{4}, Neal Katz\altaffilmark{5}, Romeel A. Dav{\'e}\altaffilmark{6,7,8}, Mark Fardal\altaffilmark{5}, Piero Madau\altaffilmark{9}, Charles Danforth\altaffilmark{3}, Amanda B. Ford\altaffilmark{10}, Molly S. Peeples\altaffilmark{11}, Joseph McEwen\altaffilmark{2}}

\altaffiltext{1}{Observatories of the Carnegie Institution of Washington,
  813 Santa Barbara Street, Pasadena, CA 91101}
\altaffiltext{2}{Astronomy Department and CCAPP, Ohio State University, Columbus, OH 43210, USA} 
\altaffiltext{3}{Astronomy Department, University of Colorado, Boulder, CO}
\altaffiltext{4}{Dipartimento di Scienza e alta Tecnologia, Universit{\`a} dell'Insubria, via Valleggio 11, 22100 Como, Italy}
\altaffiltext{5}{Astronomy Department, University of Massachusetts, Amherst, MA 01003, USA}
\altaffiltext{6}{ University of the Western Cape, Bellville, Cape Town 7535, South Africa}
\altaffiltext{7}{South African Astronomical Observatories, Observatory, Cape Town 7925, South Africa}
\altaffiltext{8}{African Institute for Mathematical Sciences, Muizenberg, Cape Town 7945, South Africa}
\altaffiltext{9}{Department of Astronomy \& Astrophysics, Univerisity of California, 1156 High Street, Santa Cruz, CA 95064}
\altaffiltext{10}{Astronomy Department, University of Arizona, Tucson, AZ 85721, USA}
\altaffiltext{11}{Space Telescope Science Institute, Baltimore, MD USA}
  
  \begin{abstract}

We examine the statistics of the low-redshift \lya\ forest from smoothed particle hydrodynamic simulations in light of recent improvements in the estimated evolution of the cosmic ultraviolet background (UVB) and recent observations from the Cosmic Origins Spectrograph (COS).   We find that the value of the metagalactic photoionization rate (\gh) required by our simulations to match the observed properties of the low-redshift \lya\ forest is a factor of 5 larger than the value predicted by state-of-the art models for the evolution of this quantity.  This mismatch in \gh\ results in the mean flux decrement of the \lya\ forest being underpredicted by at least a factor of 2 (a 10$\sigma$ discrepancy with observations) and a column density distribution of \lya\ forest absorbers systematically and significantly elevated compared to observations over nearly two decades in column density.  We examine potential resolutions to this mismatch and find that either conventional sources of ionizing photons (galaxies and quasars) must be significantly elevated relative to current observational estimates or our theoretical understanding of the low-redshift universe is in need of substantial revision.
 \end{abstract}
  
 \section{Introduction}

By virtue of its physical and chemical simplicity, the intergalactic medium (IGM) serves as an exquisite calorimeter, recording the instantaneous ionizing emissivity and heat produced by cosmic sources at each epoch.  At $z<6$ the IGM is highly ionized (Gunn \& Peterson 1965), with a fluctuating residue of neutral hydrogen: the \lya\ forest \citep{lynds71}.   After nearly two decades, the \lya\ forest remains the most well-understood and robust prediction of cosmological hydrodynamic simulations \citep[e.g][]{cen94, zhang95, miralda96, hernquist96, rauch97}.  This robustness arises because the \lya\ forest is dominated by gas at moderate overdensity;  gas that traces the dark matter (with only mild impact by pressure forces) and whose temperature is governed by the simple processes of photoionization heating and adiabatic cooling \citep[e.g.][]{weinberg98, peeples10a}.  It is this simplicity that makes the calorimeter reliable: IGM models suggest the neutral fraction of gas at the cosmic mean density at $z\sim 3$ is $n_{HI}/n_{H} \sim 10^{-5.5}$ \citep[e.g.][]{kollmeier03}, and this low neutral fraction {\it must} be maintained by the background of photo-ionizing radiation produced by quasars and star-forming galaxies \citep{miralda90,hm96}, possibly augmented by other undiscovered sources.

Determining the intensity of the ultra-violet background (UVB) --- specifically the hydrogen photoionization rate, \gh, is non-trivial.  Because of its low surface brightness, the metagalactic UVB is not directly measured but is inferred through multiple independent channels, such as the quasar proximity effect \citep[e.g.][]{bechtold87} or the low surface brightness emission from the outskirts of galactic disks \citep[e.g.][]{adams11}.   The measurements are intrinsically difficult and subject to significant uncertainties and challenges (e.g. anisotropic QSO radiation, uncertain local gas densities).   Alternatively, one can {\it predict} the intensity and spectrum of the UVB by synthesizing measurements of all possible sources of ionizing flux and the absorption and re-emission of UV radiation by the IGM and high column-density absorbers \citep[][hereafter, HM01]{hm96,hm01}.   While this procedure relies on a host of observational inputs, the most uncertain is the fraction $\fesc$ of ionizing photons that escape from star-forming galaxies.  Direct (and difficult) ionizing continuum measurements suggest  that $\fesc \sim10\%$ at $z \sim 3-4$ (e.g. Shapley et al. 2006, Vanzella et al. 2010) and is substantially lower at $z < 1$ \citep[e.g.][]{bridge10, barger13}.

Exploiting our theoretical understanding of the IGM provides a third avenue for probing \gh.  By forcing a match between the (more easily observed and well understood) opacity of the \lya\ forest and that from a theoretical IGM (typically taken directly from simulations) we have an independent determination that can be compared with both indirect measurements and the predicted UVB.  There has historically been excellent agreement between the predicted UVB and the value inferred from the mean opacity.  It was precisely the comparison between the predicted and observed forest opacity that provided strong arguments (now confirmed) for a  ``high'' baryon density and low associated deuterium abundance \citep{rauch97,weinberg97}.

\begin{figure}[H]
\plotone{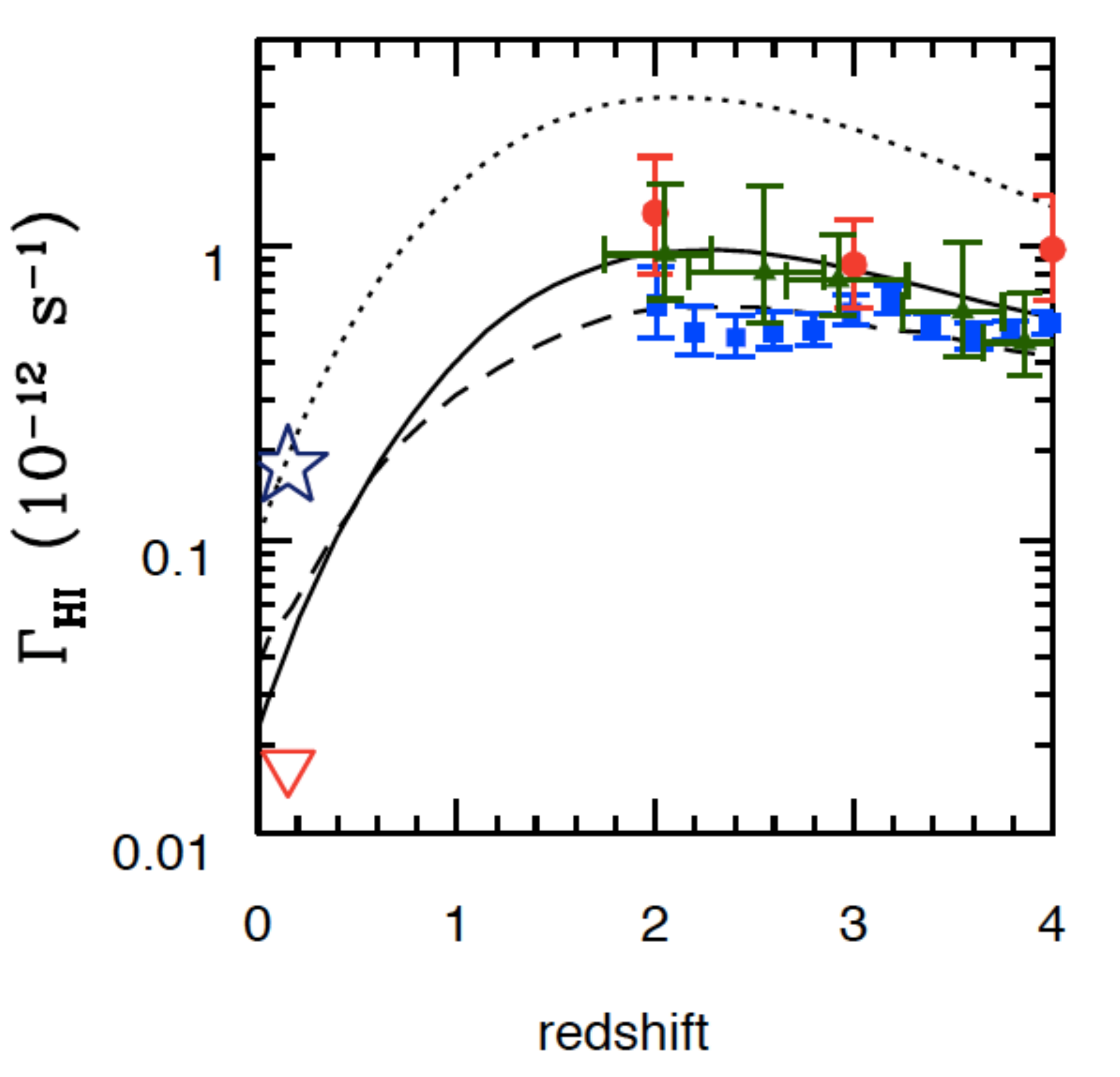}
\caption{ The photoionization rate as a function of redshift for the HM12
UVB (solid) compared to observational constraints at $z=2-4$ (circles \citet{bolton07}, triangles \citet{becker07} and squares \citet{faucher-giguere08}) and the value we infer
from our \lya\ forest modeling at $z=0.1$ (open star).  The red triangle shows the low-redshift upper limit inferred by Adams et al. (2011)
from non-detection of H$\alpha$ emission in the galaxy UGC 7321.
The dashed line shows an alternative UVB model from \cite{faucher-giguere09}.  The dotted line shows a model, discussed in \S 4.1, with a constant galaxy escape fraction $\fesc=15\%$.}
\label{fig:background}
\end{figure}

In this paper, we demonstrate that this excellent agreement no longer holds at low redshift.  Specifically, we will show a factor of $\sim 5$ discrepancy between the \gh\ predicted by the most sophisticated model of UVB evolution (Haardt and Madau 2012; hereafter HM12) and the value required to reproduce observed properties of the \lya\ forest.  We show in Figure~\ref{fig:background} the predicted \gh\ from HM12 (black solid line) compared to observational determinations.   The dashed line shows an independent model of the UVB from \cite{faucher-giguere09}, which overall is quite similar to that of HM12.  The large open star, is the value reported here.

We take advantage of new measurements of the column density distribution (CDD) of
the low-redshift \lya\ forest from Cosmic Origins Spectrograph (COS)
observations \citep{danforth14} to determine \gh\ by comparison with cosmological simulations of the IGM.  We will further show that our conclusions would be very similar if we instead use the mean decrement as our observational measure.

After describing the cosmological simulation that we use (\S 2)
and inferring the value of \gh\ required to match the observed CDD (\S 3),
we discuss (in \S 4) possible resolutions to the discrepancy between our inferred \gh\ and the value predicted by HM12, the ``photon underproduction crisis'' (PUC) of our title.
None of these resolutions alone appears entirely satisfactory.
The most exciting possibility is that this discrepancy is probing exotic sources of ionizing photons or novel heating mechanisms in the diffuse IGM operating far above the usual theoretical expectations.

 \section{Simulations and Artificial Spectra}
 \label{sec:sims}
 The simulations in this paper are performed using a modified version of the Gadget2.0 SPH+nbody code \citep{gingold77, lucy77, springel02, springel05, oppenheimer08}.  We include radiative cooling from primordial composition gas and metals assuming ionization equilibrium.  Our main simulation is performed within a box of 50Mpc/h on a side (co-moving) with $2 \times 576^3$ particles and LCDM cosmological parameters  $\Omega_{m} = 0.25$, $\Omega_{\Lambda}=0.75$, $\Omega_b=0.044$, $h=0.70$.  Our principal results are obtained from the $z=0.1$ output of this simulation. 

We stress at the outset that the results presented here should be robust to the adopted simulation code.  At the physical scales and conditions of the \lya\ forest, numerical issues of co-existing multi-phase media and under-or-over resolved hydrodynamic instabilities do not play a significant role.  While our simulation adopts the momentum-driven wind (``vzw") formalism (described in detail in \citet{oppenheimer08}), and these winds can significantly heat the IGM close to galaxies, the overall \lya\ forest properties of our simulations are largely insensitive to the adopted galactic wind prescription \citep[e.g.][]{kollmeier06, mcdonald06, dave10}.

 We extract 2500 spectra from our simulation boxes along a fixed grid using the {\sc Specexbin} software package (Oppenheimer \& Dav{\'e} 2006).  At each pixel position in the simulated sightline, the physical properties of the gas are computed (density, temperature, velocity and metallicity) by considering the contribution of each overlapping SPH particle.  The ``physical space" spectra are converted to redshift space by incorporating the gas peculiar velocity and thermal motions as well as the Hubble flow along the line of sight.  The spectra are finally convolved with the instrumental profile of the Cosmic Origins Spectrograph aboard the {\it Hubble Space Telescope}.    With the spectra in hand, we analyze the individual absorption systems in a way to mimic, as closely as possible, the corresponding observational procedure \citep{dave97}.

\section{Results}
\label{sec:results}

\begin{figure}[H]
\plotone{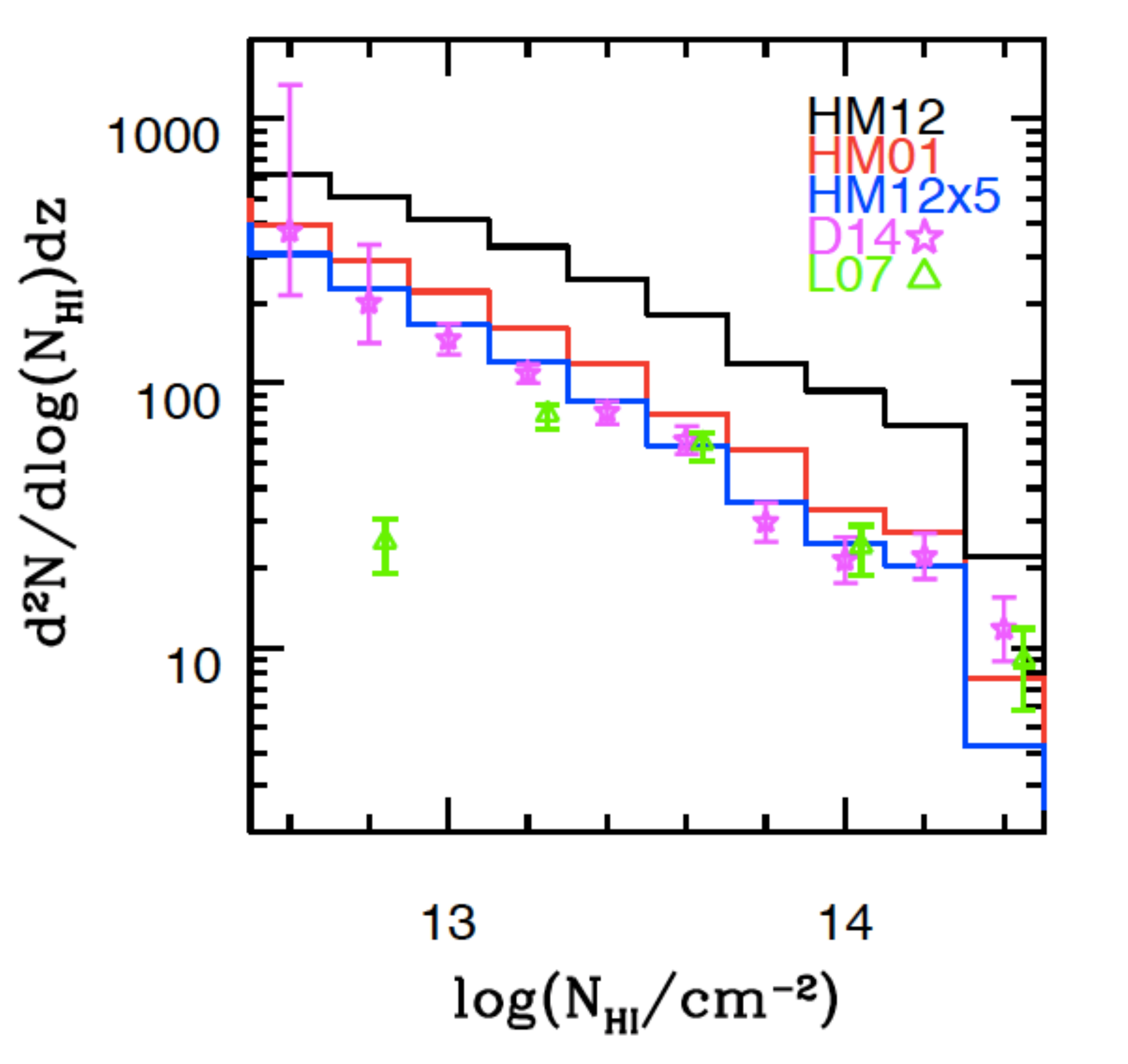}
\caption{Column density distribution in the low-redshift IGM.  Black (red) line shows the column density distribution determined from simulations adopting the HM12 (HM01) UVB estimates.  The blue data points shown are from COS observations from Danforth et al. 2014, while green symbols show the data from Lehner et al. 2007.  The blue lines show a model in which HM12 is ``boosted" by a constant factor of 5 (HM12 $\times$ 5).}
\label{fig:cdd}
\end{figure}

Figure~\ref{fig:cdd} shows the CDD, defined here to be
the mean number of absorbers per logarithmic interval of column density
per unit redshift pathlength, for the simulated \lya\ forest created
with the HM12 and HM01 backgrounds.  The COS CDD measurements from \cite{danforth14} are shown as magenta symbols.  
Earlier CDD measurements by \cite{lehner07} (green symbols) are in excellent agreement at column densities above $10^{13}\cm^{-2}$ 
and are likely affected by incompleteness in the lowest column density bin.  The simulated IGM is a thicker beast with the HM12 UVB determination, overpredicting the observed
CDD by a factor of $\approx 3.3$ over the column density range
$10^{13}-10^{14}\cm^{-2}$.  With the HM01 background, the simulated
CDD is slightly but consistently above the COS measurements.

For a highly ionized system in photoionization equilibrium, the
neutral column density is inversely proportional to the photoionization
rate, \Nhi$ \propto 1/$\gh.  The slope of the simulated and observed
CDDs in Figure~\ref{fig:cdd} is approximately $N_{\rm HI}^{-0.75}$, so
reducing the amplitude of the CDD by a factor of 3.3 requires lowering the column densities of individual systems by a factor
of $3.3^{1/0.75} \approx 5$.  The blue histogram in Figure~\ref{fig:cdd}
shows the CDD computed from the simulation with the HM12 UVB increased in amplitude by a factor of five, and the agreement
with the \cite{danforth14} data is now very good.  There is some
tension in the highest column density bin, where line saturation
effects are beginning to become important and details of fitting
algorithms, observational noise, and spectral resolution may play
a larger role.

The mean flux decrement for the simulated spectra is
$\langle D \rangle = 0.05$, 0.024, and 0.018 for the HM12,
HM01, and boosted HM12 backgrounds respectively.
These can be compared to the value of $\langle D \rangle = 0.020\pm 0.003$
found by \cite{kirkman07} from observed \lya\ forest spectra at
low redshift.  The mean decrement results lead to exactly the same
conclusion as the CDD analysis, requiring a factor of $\approx 5$
boost in the amplitude of \gh\ relative to the HM12 prediction.
The CDD comparison has the virtue of focusing on systems that
are the most physically simple\footnote{While the mean decrement is dominated by low column density \lya\ forest
systems, there is a non-negligible contribution from high column density systems associated with galaxies that can, in principle, complicate both the observational and theoretical analysis} (i.e., moderate overdensity) and the most straightforward to measure observationally,
but the similarity of the results demonstrates the robustness
of the conclusion.

\section{Discussion} \label{sec:discussion}

The large mismatch between the low-redshift photoionization rate predicted by
HM12 and inferred from matching the observed CDD ``challenges"\footnote{Significantly.  Hence "Photon Underproduction {\it Crisis}"} our current understanding of the sources of the UVB, the physical state
of the IGM, or both.  We now discuss a number of possible resolutions to this discrepancy.

\subsection{Galaxy Escape Fraction}
\label{sec:escape}

From the point of view of the low-redshift \lya\ forest, the most important change between HM01 and HM12 is a different model for the escape fraction of ionizing photons from galaxies, $\fesc$.\footnote{Defined here to
represent the average fraction of 1-4 Ryd photons that
escape their host galaxies.}
To simultaneously match the high escape fractions required
at $z > 5$ to explain reionization and the much lower $\fesc$ inferred from Lyman continuum observations of galaxies at $z \sim 3$ (typically a few percent, e.g., \citealt{shapley06,boutsia}), HM12 adopt an evolving mean escape fraction $\fesc = 1.8 \times 10^{-4} (1+z)^{3.4}$.
As a result, the contribution of galaxies to \gh\ is modest
at $z = 3$ and negligible at $z=0$, while in HM01 the
galaxy and quasar contributions are comparable at both redshifts.
From calculations with HM12's
CUBA code, we find that producing our inferred low-redshift \gh\
while keeping other aspects of the HM12 model fixed requires
an escape fraction $\fesc \approx 15\%$, as shown by the dotted
curve in Figure~\ref{fig:background}.  This simple model significantly overshoots (factor $\sim 3$) the background at $z = 2-4$.  Therefore, resolving the PUC with ionizing
radiation from galaxies would require $\fesc$ to evolve non-monotonically between $z=6$ and $z=0$.  Most seriously, the required $\fesc$ is
incompatible with most direct searches for Lyman continuum
radiation from star-forming galaxies at $z = 0-1.5$  (e.g., \citealt{bridge10, barger13}).

\subsection{Quasar Emissivity} \label{sec:quasar}

The low-redshift quasar emissivity in HM12 is about a factor of two
below that in HM01, a consequence of changing the adopted mean
quasar SED and the adopted luminosity function (primarily the latter).  The quasar SED in HM12 is harder than in HM01, bringing it into significantly
better (although not perfect) agreement with the most recent
measurements of the SED shape \citep{shull12}.
Recent estimates of the evolving quasar luminosity function also imply a somewhat reduced emissivity at low redshift
(e.g., \citealt{hopkins07,cowie09}). At the factor of two level the estimated contribution of quasars to the low-redshift UVB has remained stable for nearly
two decades; for example \cite{shull99} infer
\gh$(z=0) = 3.2\pm 1.2 \times 10^{-14} \sec^{-1}$, which is compatible with
the HM12 value of $2.3 \times 10^{-14} \sec^{-1}$.
It therefore seems unlikely that this contribution could be increased
by a factor of five relative to the HM12 value, or even by
the factor of two that would get back to HM01's estimate of
the quasar emissivity.  Corrections of this magnitude would
require a dominant contribution from systems that
have largely escaped the existing census of AGN.  In addition to being heretofore invisible, these systems would necessarily have substantially different SEDs
from standard AGN so that they could dominate the UVB without
overproducing the directly observed X-ray background.

\subsection{Mean Free Path}

At high redshift, the intensity of the UVB is controlled in part by the mean free path of ionizing photons, which sets the
horizon over which a given source's Lyman continuum radiation
can influence the IGM.  HM01 and HM12 compute
the radiative transfer of UVB photons using observational estimates
of the frequency of high column density \lya\ absorbers
(\Nhi$ \ga 10^{15}\cm^{-2}$).  While these estimates remain
somewhat uncertain, the mean free path becomes large at $z < 1$
in any case, so that the horizon is set by cosmological redshifting
rather than by absorption.  As a test, we have run CUBA in a case where we double the mean free path
 at $z < 1$, a change already inconsistent with observations \citep{stengler-larrea95,ribaudo11}, and find that \gh$(z=0)$ only increases by a factor of 1.4.  

\subsection{Extra Heating of the IGM}
\label{sec:heating}
It is possible that there is a source of IGM heating that is not accounted for by our simulations.  In conventional models, the temperature-density relation of the diffuse IGM is determined by the balance between heating by photoionization and adiabatic cooling by the expansion of the
universe \citep[e.g.][]{katz96}.  For temperatures $T \la 2\times 10^5\,{\rm K}$, the hydrogen recombination coefficient scales as $T^{-0.7}$, so a hotter IGM produces a thinner \lya\ forest.
The HI column density of a given system scales as $T^{-0.7}$\gh$^{-1}$, so a factor of five increase in \gh\ achieves the same effect as a factor of  $5^{1/0.7}\sim 10$ increase in temperature.

To examine the amount of heating required to resolve this discrepancy, we have carried out the simple experiment of boosting the temperature of all SPH particles in the simulation by
a factor of four before extracting \lya\ forest spectra with the HM12 UVB,
and we find that the resulting CDD is similar to but slightly
above that of the HM01 model shown in Figure~\ref{fig:cdd}, as expected by comparing $4^{0.7} = 2.6$ to the factor of 3.7 difference in \gh\ between HM01 and HM12.

A harder ionizing background spectrum leads to a hotter IGM
temperature because the average residual energies of photoelectrons are
higher.  However, producing even a factor of four increase in temperature would require an implausibly hard UVB spectral shape,
so the extra heating solution would require some mechanism other
than photoionization.   \cite{broderick12} have proposed one such
mechanism, powered by TeV $\gamma$-ray emission from blazars.  These high energy gamma rays annihilate and pair-produce through interactions with extragalactic background photons.  In principle, this process can drive a plasma instability, which locally dissipates the (high) energy of the produced pair, thereby heating the low-density IGM.   This heating results in an {\it inverted} temperature-density relation in the forest because it deposits more energy per particle (Puchwein et al. 2012).

\begin{figure}[H] \plotone{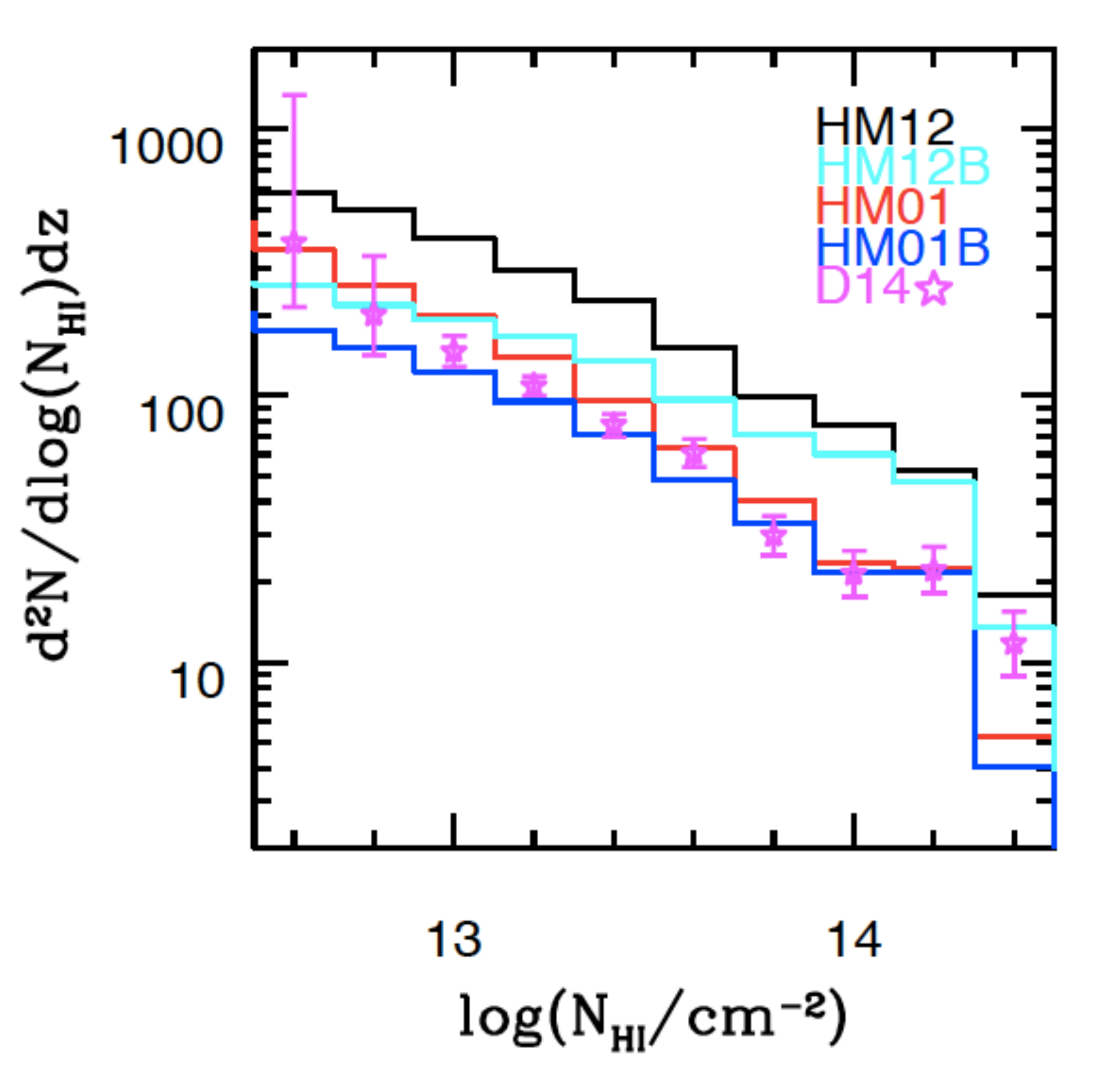}
\caption{Effect of blazars on the low-redshift \lya\ forest statistics. Black, cyan,
red, and blue lines correspond to simulations performed with the HM12
background, HM12 background plus blazars, HM01 background, and HM01
background plus blazars, respectively.  } \label{fig:blazars} \end{figure}

To investigate the potential impact of blazar heating, we
have performed a suite of simulations implementing the ``intermediate"
prescription for blazar heating proposed by \cite{puchwein12},
and confirmed that we reproduce their results for the $\rho-T$ relation at $z=3$ and $z=0$.
This simulation was performed at slightly lower resolution, with
$2\times 384^3$ particles in a $48\hmpc$ box, and we ran
a matched simulation without blazar heating for comparison.  The blazar heating has the anticipated effect of thinning out the \lya\ forest at low redshift for both the HM12 and HM01 UVB models.  The shallower CDD for the blazar
simulation is because of the inverted $\rho-T$ relation, which reduces the neutral
fraction in the lower density gas responsible for lower
column density absorbers.  This highlights the potential to use the low-redshift \lya\ forest CDD to probe the impact of blazar contributions to the heating of the low-density IGM.  However, blazar heating is clearly
insufficient to reconcile the HM12 background with the
observed CDD.  If we instead apply the HM01 background to
the blazar simulation we get reasonably good agreement
with the \cite{danforth14} CDD, but most of this improvement comes from the lower $\ghi$ of HM01.

\newpage

\subsection{Density Structure of the IGM}
\label{sec:simbad}
It is possible that our cosmological simulations are predicting the wrong density structure of the diffuse IGM owing to either cosmological or numerical inaccuracies.  We think both of these are extremely unlikely.  At our simulation resolution, the properties of the forest are well-converged and largely insensitive to e.g. wind model, feedback prescriptions, spectral resolution and noise characteristics \citep{dave10, peeples10b}.  A cosmological solution would require making the low redshift
IGM much {\it smoother} than LCDM simulations predict -- shifting absorption systems systematically to lower columns or erasing them entirely.  It is possible that cosmological models including warm dark matter or a small scale cutoff in the primordial power spectrum would go in this direction, but we are skeptical that any such change could resolve the PUC while maintaining a
good match to the observed \lya\ forest.  With our preferred
UVB intensity, on the other hand, LCDM simulations match
observed \lya\ forest and metal-line absorption statistics
over a wide range of redshifts.

\subsection{New Sources of Ionizing Photons}
\label{sec:newsources}

The most exciting interpretation of the PUC
is that it has revealed the presence of previously unrecognized
(and dominant) sources of ionizing photons in the low redshift universe.
A population of low luminosity or hidden AGN could be such a source, though this would require a major revision to our understanding of
the AGN luminosity function.  Lyman continuum searches have mostly
focused on rapidly star-forming galaxies, and it is possible that
a different population, such as early-type galaxies producing
UV radiation from core helium burning or post-AGB stars,
dominates the production of ionizing photons.
More exotically, the ``missing'' photons could be coming from
decaying or annihilating dark matter particles in the dense cores of halos and subhalos.
Producing the implied energy density of the low redshift UVB
would only require the decay of $\sim 10^{-10}$ of the dark matter density over a Hubble time, so this solution would
not significantly alter cosmological parameters, but it would
have profound implications for dark matter properties.

Further studies of H$\alpha$ emission from nearby galaxies are an important
area for future work, as confirmation of the \cite{adams11} limit on $\ghi$ would rule out {\it all} solutions involving additional
ionization, including those in \S\S 4.1-4.3.

\section{Conclusions}

The factor of 5 discrepancy between the value of \gh\ required to match cosmological models of the $z=0$ IGM to the observed mean decrement and CDD and that predicted by state-of-the art models for the evolution of the extragalactic UVB (HM12) highlights a significant gap in our current understanding of the sources of the UV background or the structure of the IGM, or both. We have discussed a number of possible resolutions, no one of which appears satisfactory.   The {\it least} radical solution is to increase the mean $\fesc$ at low-redshift such that galaxies dominate the emissivity {\it and} simultaneously boost the quasar emissivity, though both of these changes oppose our current understanding of these sources.  For the undaunted,  extra photons from decaying dark matter or a drastic change to the physical structure of the IGM as predicted by LCDM may also be the resolution to the photon underproduction crisis.

 \section*{Acknowledgements}
   We thank Rik Williams and Andy Gould for helpful discussions and comments.  Support for this work was provided by the NSF through grant OIA-1124453 and by NASA through grants NNX12AF87G, NNX10AJ95G, and HST-AR-13262.  Computing resources used for this work were made possible by a grant from the Ahmanson Foundation.

  \end{document}